\documentclass[oldversion]{aa}
\usepackage{epsfig}
\usepackage{natbib}
\usepackage{lscape}
\usepackage{wasysym}
\bibpunct{(}{)}{;}{a}{}{,}
\begin{document}

\title{The HARPS search for southern extrasolar planets\thanks{Based on 
observations collected at the La Silla Parana Observatory,
ESO (Chile) with the HARPS spectrograph at the 3.6-m telescope (ESO runs ID 
72.C-0488 and 082.C-0212). Tables 3, 4, and 5 (with the radial-velocities) are only available 
in electronic form at the CDS via anonymous ftp to cdsarc.u-strasbg.fr (130.79.128.5)
or via http://cdsweb.u-strasbg.fr/cgi-bin/qcat?J/A+A/ }}

\subtitle{XXI. Three new giant planets orbiting the metal-poor stars HD5388, HD181720, and HD190984}

\author{
  N.C. Santos\inst{1} \and
  M. Mayor\inst{2} \and
  W. Benz\inst{3} \and
  F. Bouchy\inst{4} \and
  P. Figueira\inst{2} \and
  G. Lo Curto\inst{5} \and
  C. Lovis\inst{2} \and	
  C. Melo\inst{5} \and
  C. Moutou\inst{6} \and
  D. Naef\inst{2} \and
  F. Pepe\inst{2} \and
  D. Queloz\inst{2} \and
  S. G. Sousa\inst{1} \and
  S. Udry\inst{2}
  }

\institute{
    Centro de Astrof{\'\i}sica, Universidade do Porto, Rua das Estrelas, 
    4150-762 Porto, Portugal
    \and
    Observatoire de Gen\`eve, Universit\'e de Gen\`eve, 51 ch. des Maillettes, 1290 Sauverny, Switzerland
    \and
    Physikalisches Institut Universit\"at Bern, Sidlerstrasse 5, 3012 Bern, Switzerland
    \and
    Institut d'Astrophysique de Paris, UMR7095 CNRS, Universit\'e Pierre \& Marie Curie, 98bis Bd Arago, 75014 Paris, France
    \and
    ESO - European Southern Observatory, Karl-Schwarzschild-Strasse 3, 85748 Garching bei M\"unchen, Germany
    \and
    Laboratoire d'Astrophysique de Marseille, Traverse du Siphon, 13376 Marseille 12, France
}


\date{Received  ; accepted  }

\abstract{
We present the discovery of three new giant planets around three metal-deficient stars: HD\,5388\,b (1.96\,M\,$_{Jup}$),
HD\,181720\,b (0.37\,M\,$_{Jup}$), and HD\,190984\,b (3.1\,M\,$_{Jup}$). All the planets have moderately eccentric orbits (ranging from 0.26 to 0.57) and long orbital periods (from 777 to 4885\,days). Two of the stars (HD\,181720 and HD\,190984) were part of a program searching for
giant planets around a sample of $\sim$100\, moderately metal-poor stars, while HD\,5388 was part of the volume-limited sample of
the HARPS GTO program. Our discoveries suggest that giant planets in long period orbits are not uncommon around moderately metal-poor stars. 
 }

\authorrunning{Santos et al.}
\maketitle

\section{Introduction}

The discovery of about 400 exoplanets orbiting solar-type stars
has opened a number of questions about the origin of the newfound planets
\citep[for a review see][]{Udry-2007}. The theories of planet formation 
are thus confronted with new and fascinating problems, whose solution
may give us a new insight into the processes of planet formation and evolution. 

Two major theories have been proposed to explain the formation of the discovered giant planets. 
On the one hand, the ``traditional'' core-accretion model \citep[][]{Pollack-1996}, more recently 
improved to include the effects of planet migration and disk evolution \citep[e.g.][]{Ida-2004a,Mordasini-2009a}, 
suggests that giant planets are formed by the accretion of gas around a pre-formed metal-rich (icy) core. 
On the other hand, the so-called disk instability models advocate that giant planets can be formed
by the direct collapse of gas and dust in the proto-planetary disk \citep[][]{Boss-1997}. 

Based on recent observational results \citep[see discussion in][]{Udry-2007}
the core-accretion model is presently favoured. However, the existence of giant planets at more than 20\,AU from 
their host stars \citep[][]{Kalas-2008,Marois-2008} was recently suggested to be more easily explained by the 
disk instability model \citep[][]{DodsonRobinson-2009}. To understand which of the two processes of 
giant planet formation best explains the detected population of exoplanets we need more observational data.

The solution to this problem may include the understanding of the well known stellar metallicity-giant
planet correlation. It is known that giant planets are more easily found orbiting metal-rich 
dwarfs \citep[][]{Gonzalez-1997,Santos-2004b,Fischer-2005}. This result strongly favors the core-accretion model to explain the 
formation of the majority of the giant planets discovered so far \citep[e.g.][]{Matsuo-2007}. Still, it is also known that planet formation is not
completely inhibited around metal-poor dwarfs \citep[e.g.][]{Cochran-2007,Santos-2007}.
The small number of planets detected around metal-poor objects still prevents a clear analysis
of the low metallicity tail of the abundance distribution of planet-host stars though.

To bridge this gap, a number of projects have been 
started to search for planets in metal-poor samples \citep[][]{Sozzetti-2006b,Cochran-2007,Santos-2007}. In addition to these, large volume-limited samples of
solar neighborhood stars are adding new planets to the lists, some of them orbiting metal-poor stars.

In this paper we present the detection by two HARPS GTO programs of three new planets which orbit metal-poor stars.
One of these was discovered as part of a large volume-limited survey \citep[][]{Pepe-2004} for giant planets. 
The remaining two are the second and third planets discovered in the HARPS metal-poor sample \citep[][]{Santos-2007}. 
In Sects.\,\ref{sec:sample} and \ref{sec:parameters} we briefly present our sample as well as the parameters
of the host stars. In Sect.\,\ref{sec:rv} we present the radial-velocity measurerements and the fitted
orbital solutions. We conclude in Sect.\,\ref{sec:conclusions}, where we discuss the implications of the present findings.

\begin{table}[t!]
\caption{Stellar parameters for the stars analyzed in the present paper. }
\label{tab:parameters}
\begin{tabular}{lccc}
\hline\hline
\noalign{\smallskip}
Parameter  			& \object{HD\,5388}& \object{HD\,181720}    & \object{HD\,190984}\\
\hline
Spectral~type			& F6V		   & G1V		    & F8V\\
$m_v$				& 6.73		   & 7.84		    &8.76\\
$B-V$				& 0.500		   & 0.599		    & 0.579\\
Parallax [mas]			& 18.77$\pm$0.96   & 17.88$\pm$1.31	    & 9.8$\pm$2.6$\dagger$\\
Distance~[pc]			& 53$\pm$3	   & 56$\pm$4		    & 103$^{+37}_ {-21}\dagger$\\
M$_v$				& 3.10		   & 4.10		    & 3.71 \\
L [L$_{\odot}$]			& 4.60		   & 1.94		    & 2.69 \\
Radius [R$_{\odot}$]		& 1.91		   & 1.39		    & 1.53\\
$\log{R'_{\rm HK}}$		& $-$4.98	   & $-$5.01		    & $-$5.01\\
$v\,\sin{i}$~[km~s$^{-1}$]	& 4.2		   & 1.5		    &3.4\\
$T_{\rm eff}$~[K]  		& 6297$\pm$32	   & 5781$\pm$18	    &5988$\pm$25\\
$\log{g}$			& 4.28$\pm$0.06	   & 4.24$\pm$0.15	    &4.02$\pm$0.22\\
$\xi_{\mathrm{t}}$		& 1.57$\pm$0.04	   & 1.17$\pm$0.04	    &1.48$\pm$0.03\\
${\rm [Fe/H]}$			& $-$0.27$\pm$0.02 & $-$0.53$\pm$0.02	    & $-$0.48$\pm$0.06\\
Mass~$[M_{\odot}]$		& 1.21		   & 0.92		    & 0.91\\
\hline
\noalign{\smallskip}
\end{tabular}
\newline
$\dagger$ Derived in this paper (see text)
\end{table}

\section{Sample and observations}
\label{sec:sample}

The planets presented in this paper were discovered orbiting stars which were followed as part of two different HARPS
GTO programs. The first consists of a sample of $\sim$850 solar-type stars in a volume-limited 
complement (up to 57.5 pc) of the ``closer" CORALIE sample \citep[][]{Udry-2000}. More details about this sample
can be found in \citet[][]{Naef-2007}. The second program consists of a sample of $\sim$100 metal-poor dwarfs that was 
searched for the presence of giant planets. A description of this latter sample can be found in \citet[][]{Santos-2007}. 

Observations of all the targets were obtained during the Guaranteed Time Observations GTO \citep[][]{Mayor-2003b}. 
For HD181720 and HD\,190984, however, complementary 
observations were obtained in a separate HARPS program whose goal was to search for neptune-mass planets orbiting
moderately metal-poor stars. 

Radial-velocities were derived using the latest version of the HARPS pipeline. From each spectrum, other parameters 
of the HARPS cross-correlation function (CCF) such as the bisector inverse slope \citep[BIS --][]{Queloz-2001}
were also derived. 

Most of the radial-velocity measurements were not done in the simultaneous ThAr calibration mode. A strategy to average out the noise due to stellar 
oscillations \citep[][]{Santos-2004a} was also not used in most of the measurements, since this is not required for the detection of giant planets (the main goal of the two mentioned programs). We can thus expect the residuals around the orbital solutions to be above 1\,m\,s$^{-1}$.  This is particularly true for early-G and late-F stars (like our targets), since the
oscillation and granulation noise is stronger for these objects (Dumusque et al. 2009, in prep.).

\begin{figure}[t!]
\resizebox{8.5cm}{!}{\includegraphics{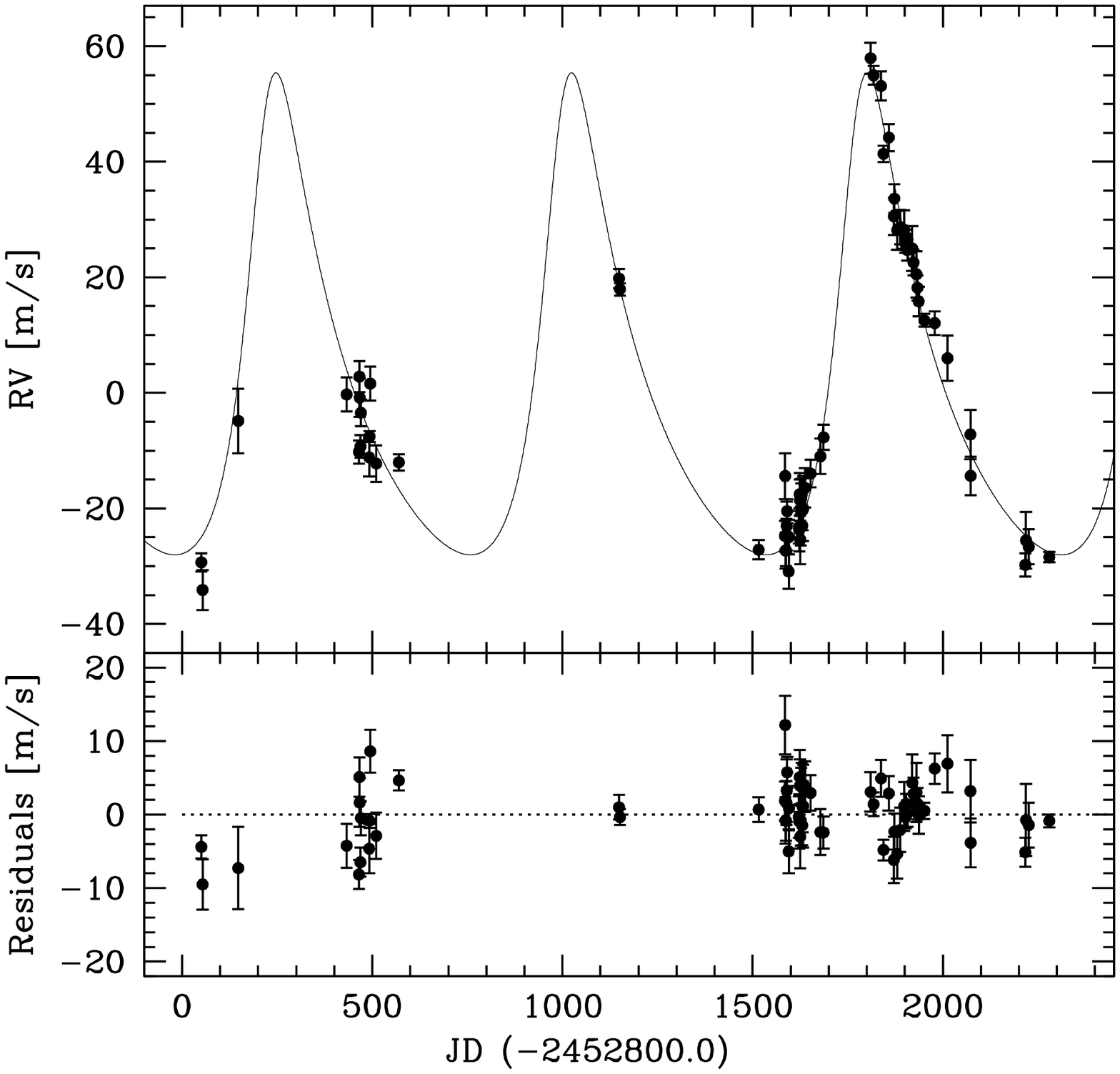}}
\resizebox{8.5cm}{!}{\includegraphics{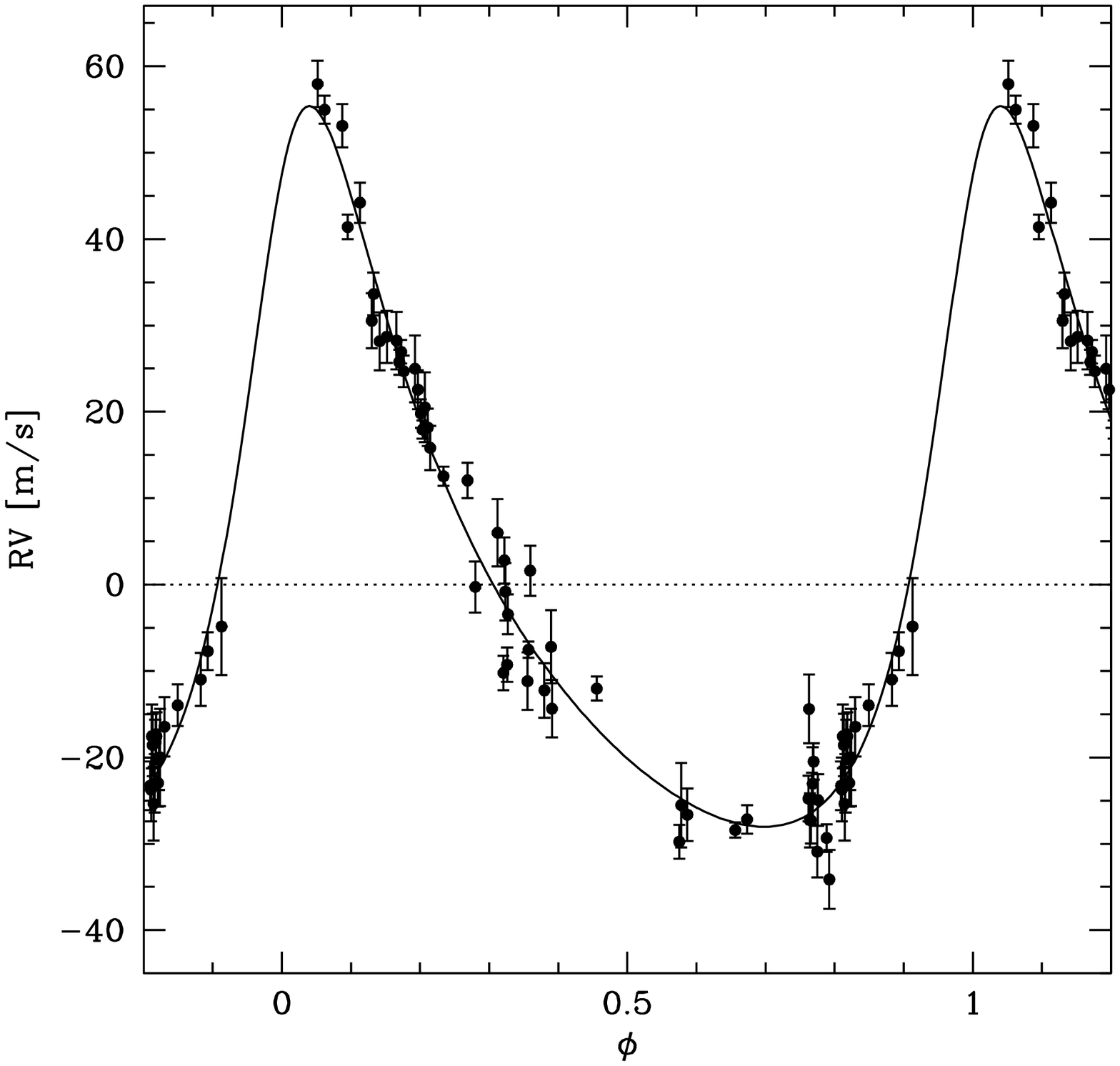}}
\caption{{\it Top}: Radial-velocity measurements of \object{HD\,5388} as a function of time, and
the best Keplerian fit to the data with a period of 777-days, eccentricity of 0.40, and semi-amplitude of 42\,m\,s$^{-1}$. The residuals of the fit are shown in the lower box. {\it Bottom}: phase-folded radial-velocity measurements of \object{HD\,5388}, and the best Keplerian fit.}
\label{fig:HD5388_rv}
\end{figure}

\begin{figure}[t!]
\resizebox{8.5cm}{!}{\includegraphics{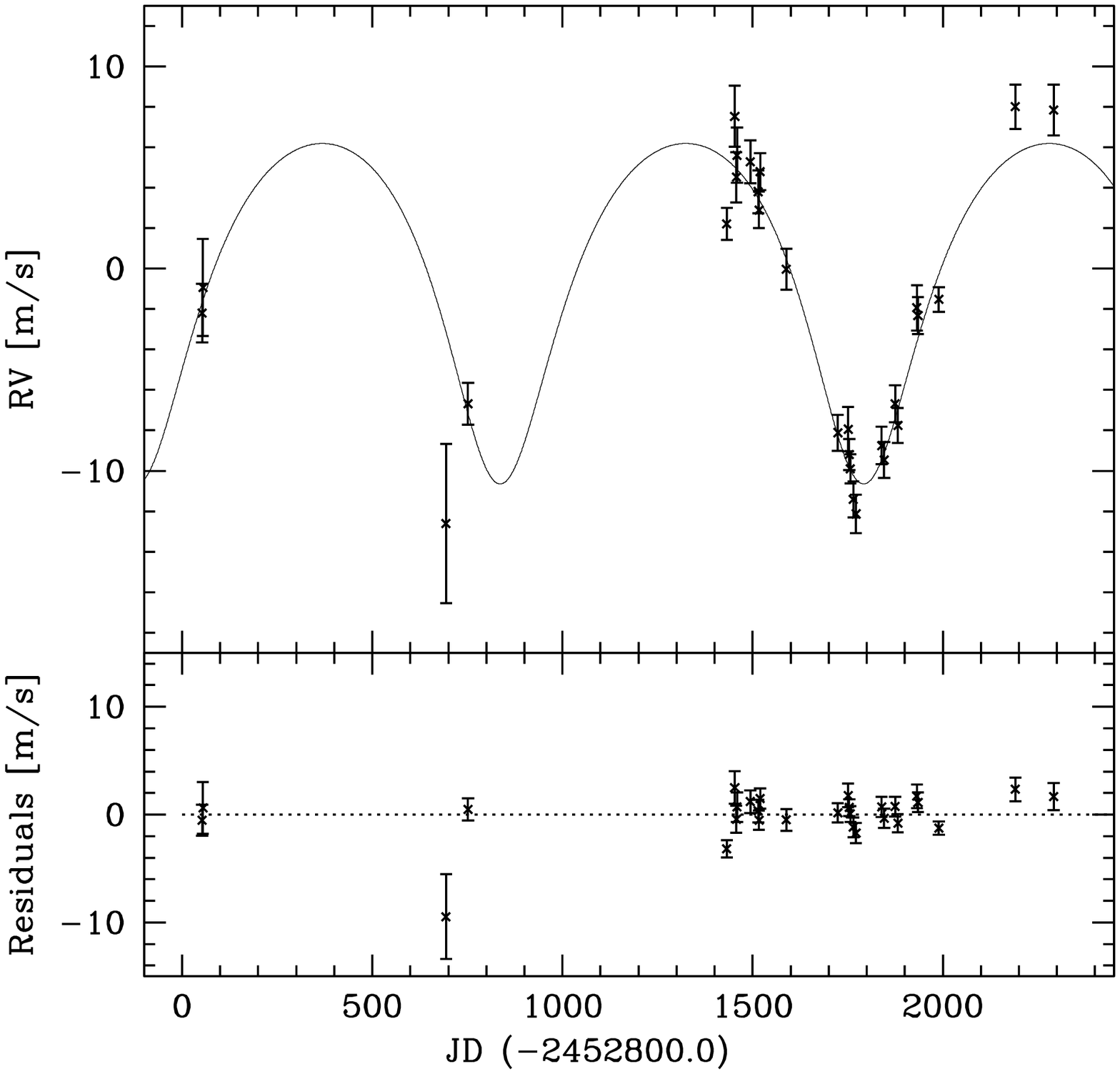}}
\resizebox{8.5cm}{!}{\includegraphics{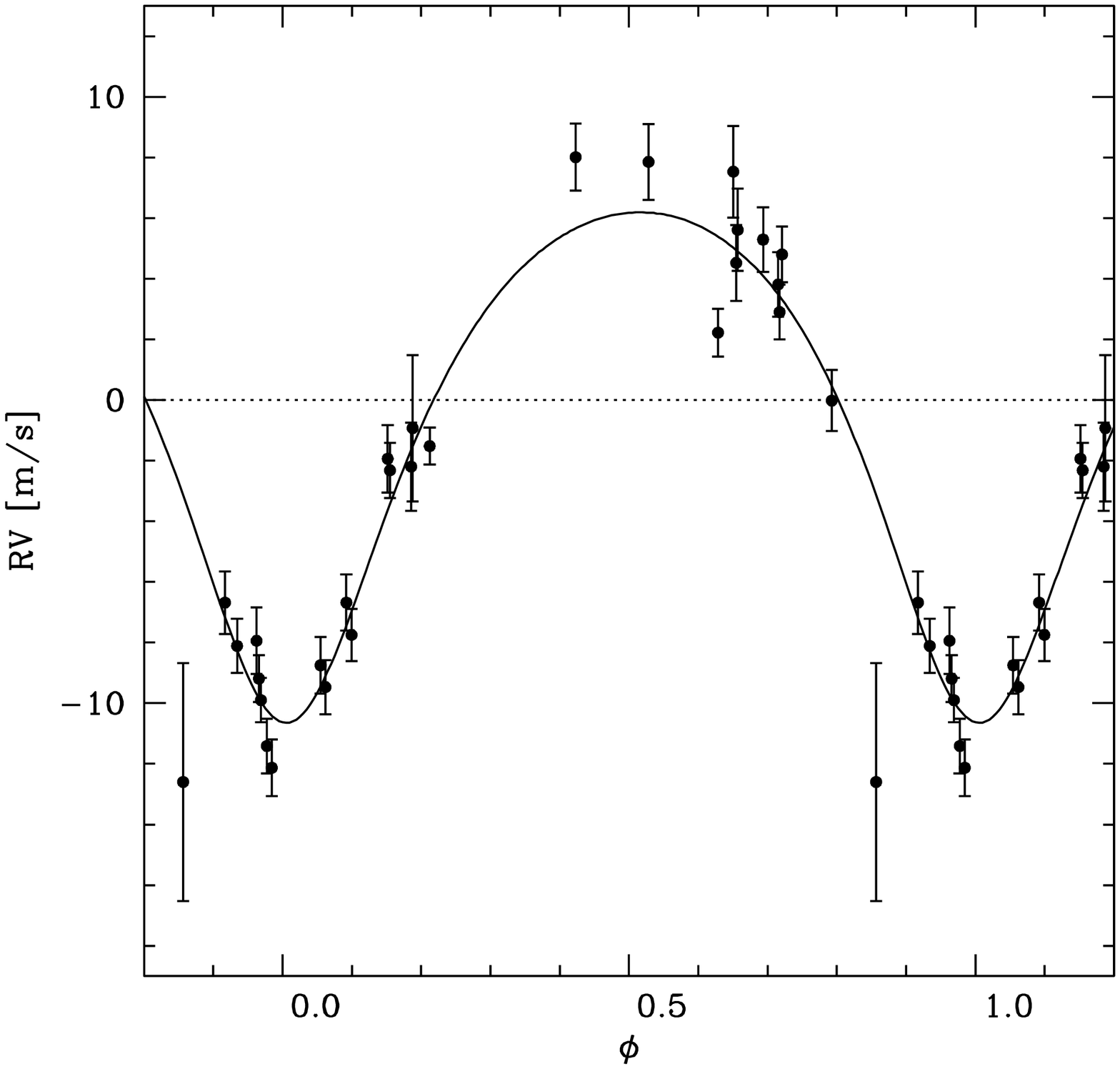}}
\caption{{\it Top}: Radial-velocity measurements of \object{HD\,181720} as a function of time, and
the best Keplerian fit to the data with a period of 956-days, eccentricity of 0.26, and semi-amplitude of 8.4\,m\,s$^{-1}$. The residuals of the fit are shown in the lower box. {\it Bottom}: phase-folded radial-velocity measurements of \object{HD\,181720}, and the best Keplerian fit.}
\label{fig:HD181720_rv}
\end{figure}

\begin{figure}[t!]
\resizebox{8.5cm}{!}{\includegraphics{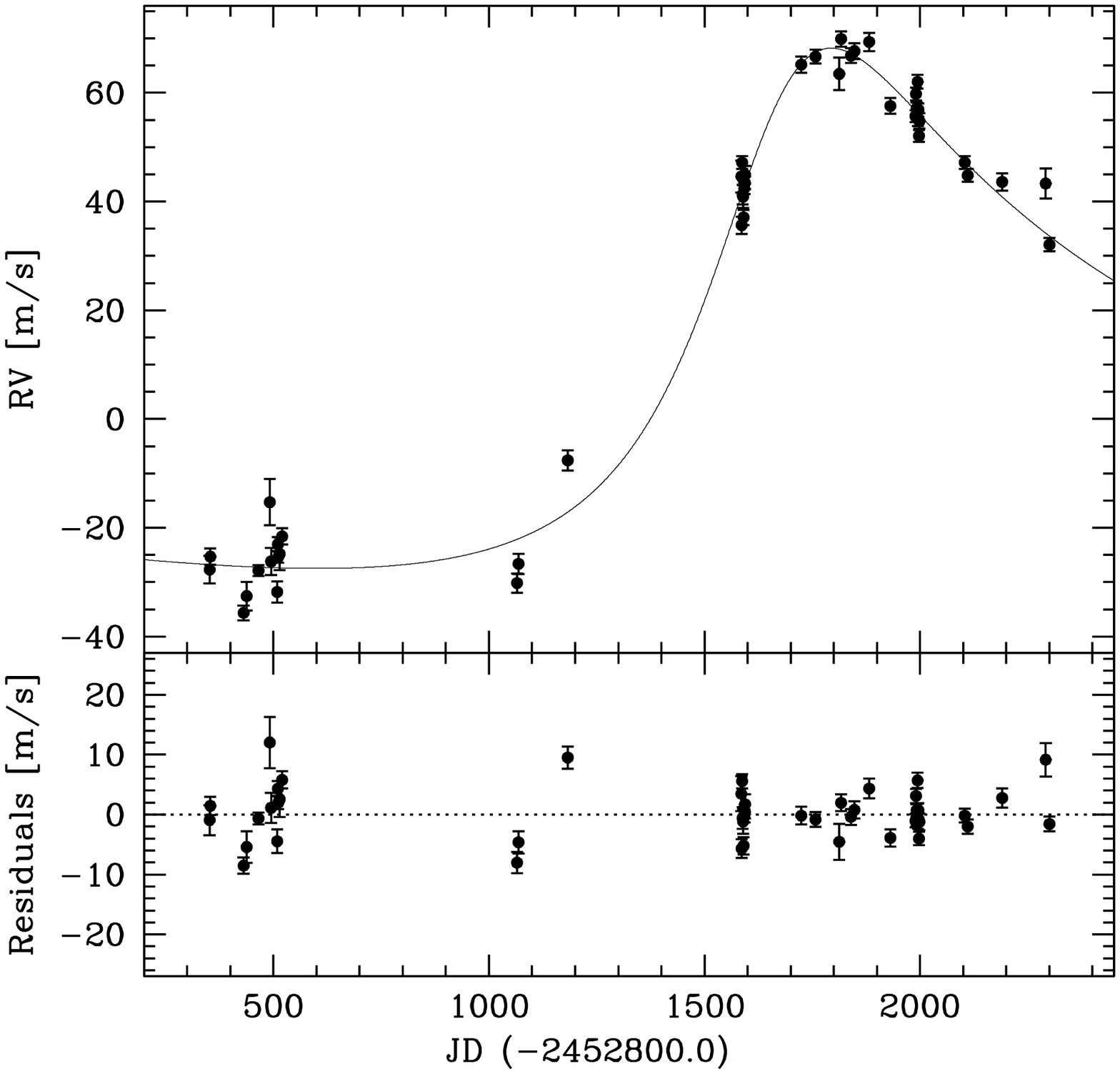}}
\resizebox{8.5cm}{!}{\includegraphics{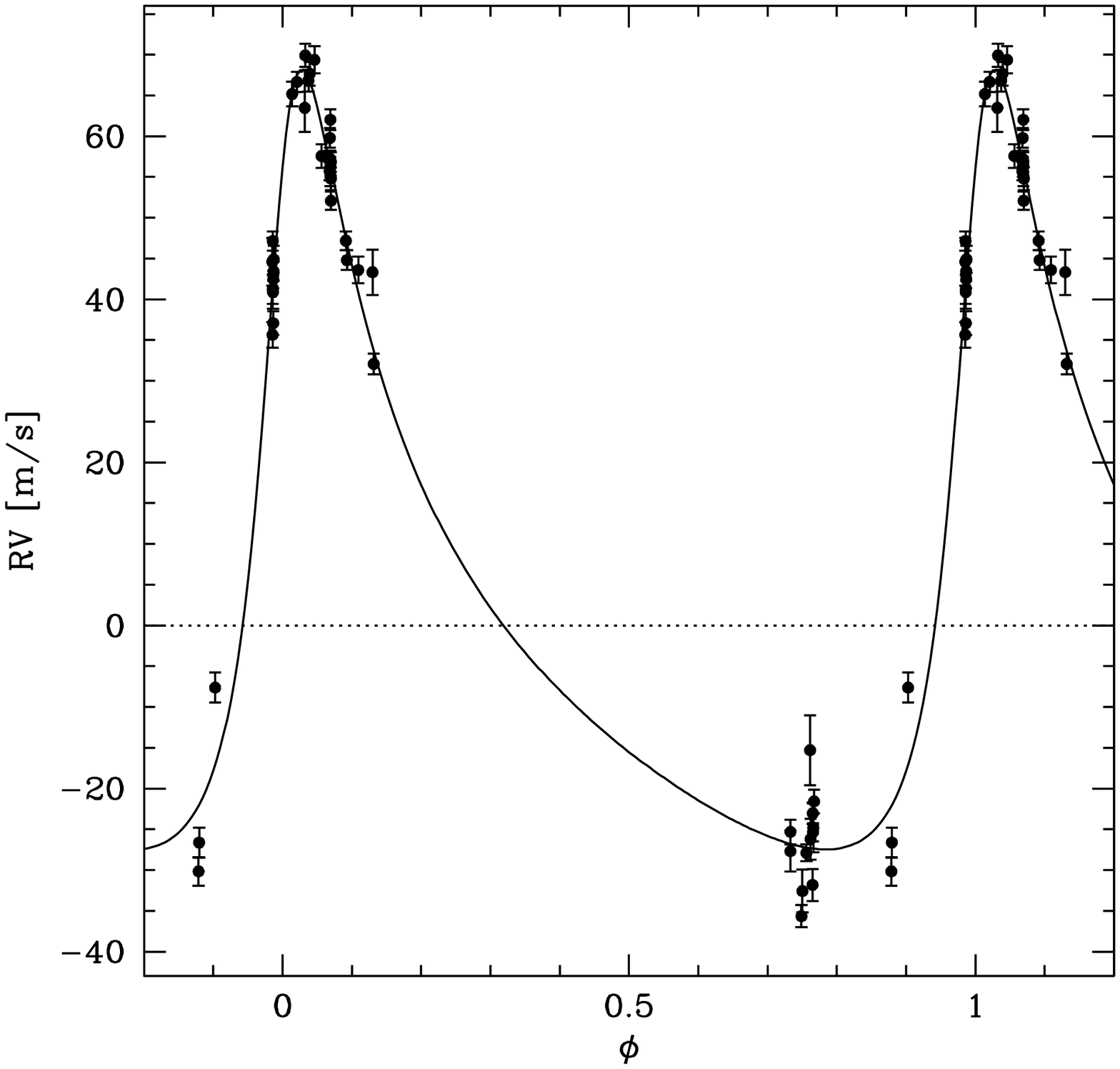}}
\caption{{\it Top}: Radial-velocity measurements of \object{HD\,190984} as a function of time, and
the best Keplerian fit to the data with a period of 4885-days, eccentricity of 0.57, and semi-amplitude of 48\,m\,s$^{-1}$. The residuals of 
the fit are shown in the lower box. {\it Bottom}: phase-folded radial-velocity measurements of \object{HD\,190984}, and the best Keplerian fit.}
\label{fig:HD190984_rv}
\end{figure}

\section{Stellar parameters}
\label{sec:parameters}

The global stellar parameters for each of the planet-host stars analyzed in the present paper are listed in Table\,\ref{tab:parameters}.

The spectral type, $m_v$, $B-V$, and parallax (and derived distance), were taken from 
the Hipparcos catalog \citep[][]{ESA-1997}, except for HD\,190984 (see discussion below). $M_{V}$, L, and the stellar radius 
were computed from the above values, using a bolometric correction from \citet[][]{Flower-1996} and the
T$_{\mathrm{eff}}$ obtained from the spectroscopy (see below).

The values for the effective temperature, surface gravity, microturbulence, and iron abundance were obtained using the high-S/N combined HARPS spectra for each target. The values were derived following the method and line-lists described in \citet[][]{Santos-2004b} and \citet[][]{Sousa-2008}. The final values, together with their errors, are 
presented in Table\,\ref{tab:parameters}. We refer the reader to these authors for more details. 

Stellar masses were also derived by interpolating the theoretical isochrones of \citet[][]{Girardi-2000}\footnote{See the web interface at http://stev.oapd.inaf.it/cgi-bin/param}, using T$_{\mathrm{eff}}$ and [Fe/H] obtained from the spectroscopy, as well as $V$ and the parallax from the Hipparcos catalog. We estimate that the uncertainties are on the order of 10\% due to the errors in the input parameters and also due to different systematic effects \citep[][]{Fernandes-2004}.

Values for the projected rotational velocity, $v\,\sin{i}$, and the stellar activity level 
(based on the \ion{Ca}{ii} H and K lines) were derived from the HARPS spectra following the general recipes 
described in \citet[][]{Santos-2000a} and \citet[][]{Santos-2002a}, respectively. The computed values show that all targets 
have a low chromospheric activity level. 

\subsection{\object{HD5388} (HIP4311)}

 \citet[][]{Masana-2006} found an effective temperature and stellar radius of 6201\,K and 1.91\,R$_{\odot}$ 
for this star in very good agreement with the values derived in the present paper (6297\,K and 1.91\,R$_{\odot}$). This value for $T_{\rm eff}$ is similar to the one derived by \citet[][6208\,K]{Nordstrom-2004} and to the value obtained using the (B-V,[Fe/H]) calibration of \citet[][6260\,K]{Sousa-2008} . The stellar metallicity derived in the present paper ($-$0.27) agrees also perfectly with the one derived by these latter authors ($-$0.25). These values are slightly above (but still compatible) with
the metallicity ($-$0.34) computed using a calibration of the HARPS Cross-Correlation Function \citep[see similar calibration for CORALIE --][]{Santos-2002a}. The derived activity value of $\log{R'_{\rm HK}}$=$-$4.98 suggests that this star is not chromospherically active.

\subsection{\object{HD181720} (HIP95262)}

The spectroscopic effective temperature obtained for this star (5781\,K) superbly agrees with the values of 5776, 5745, and 5800\,K, derived using the calibration presented in \citet[][]{Sousa-2008}, and obtained by \citet[][]{Masana-2006} and \citet[][]{Nordstrom-2004}, respectively. The metallicity of $-$0.53 is also similar to the one derived by Nordstr\"om et al. ($-$0.49), and also
to the value of $-$0.54 obtained using a calibration of the HARPS CCF. The low activity level of this star ($\log{R'_{\rm HK}}$=$-$5.01) as measured from the HARPS spectra is also supported by previous measurements of the $\log{R'_{\rm HK}}$ index of $-$5.07 and $-$5.00 derived by \citet[][]{Wright-2004} and \citet[][]{Henry-1996}, respectively.

\begin{figure*}[t!]
\resizebox{17.5cm}{!}{\includegraphics{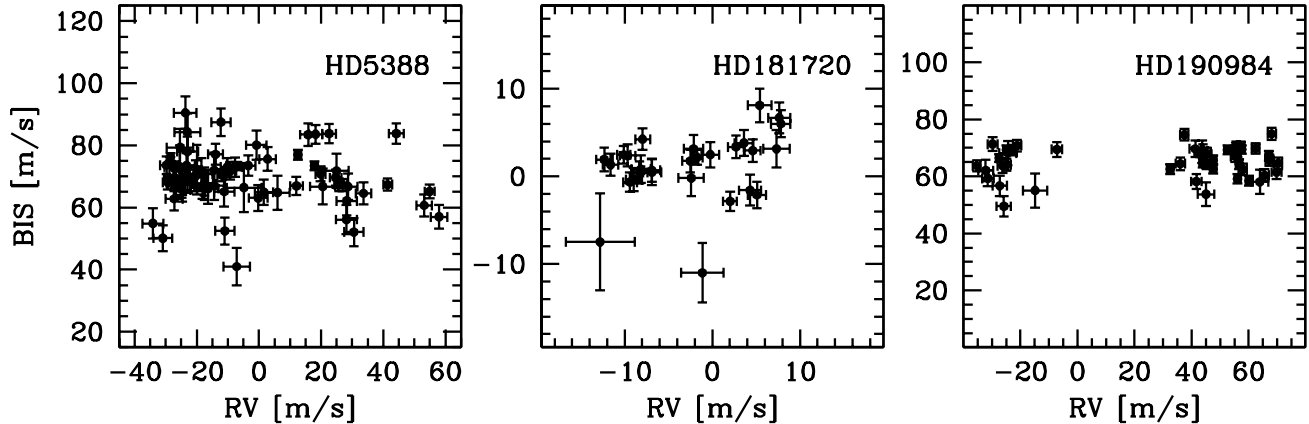}}
\caption{Bisector inverse slope (BIS) as a function of radial-velocity for the three stars. The $x$ and $y$ scales were set the same for comparison purposes}
\label{fig:bisvr}
\end{figure*}

\subsection{\object{HD190984} (HIP99496)}

The effective temperature derived for HD\,190984 (5988\,K) agrees well with the value found by \citet[][-- 5921\,K]{Masana-2006} and \citet[][5902\,K]{Nordstrom-2004}, as well as with the value obtained using the calibration of \citet[][5868\,K]{Sousa-2008}. The stellar metallicity derived by our spectroscopic analysis ($-$0.48) is also in excellent agreement with the value listed by Nordstrom et al. ([Fe/H]=$-$0.47), and slightly above the value derived using the HARPS CCF ($-$0.64).  The chromospheric activity level, as calculated using the HARPS spectra, shows that HD\,190984 is inactive. The surface gravity for this star hints that it may be slightly evolved.

The parallax value for this star listed in the Hipparcos catalog ($\pi$=5.28$\pm$1.25\,mas) has a large uncertainty. {This situation remains true if we 
take the more recent parallax value derived by 
\citet[][$\pi$=5.45$\pm$1.11]{vanLeeuwen-2007}. Making use of the Hipparcos parallax we found values for M$_{v}$ and luminosity of 2.37 and 9.3, respectively. These values seem to contradict the results from our detailed 
spectroscopic analysis.}

We have thus decided to rederive the parallax using an iterative procedure that makes use of Eq.\,1 in \citet[][]{Santos-2004b}, the relation between luminosity, radius, and parallax, and the isochrones of \citet[][]{Girardi-2000}. We first fixed the bolometric correction and the visual magnitude of the star to the values derived by the calibration of \citet[][]{Flower-1996} and the value listed in the Hipparcos catalog, respectively. An initial value for the stellar mass was also obtained using the Hipparcos parallax and $V$ magnitude and the metallicity and temperature derived from spectroscopy. Then, using this value for the mass (with an associated error of 10\%), the effective temperature and the surface gravity (and their respective errors), we took 1000 randomly selected Gaussian distributed values for these three parameters. These were then used to create a distribution for the stellar luminosities and stellar radii, which where in turn used to calculate a distribution of parallaxes. Once the new value for the parallax was derived, a new mass was obtained. Following this procedure we found a convergence after only three iterations. The final value for the parallax (see Table\,\ref{tab:parameters}) is significantly different from the value listed in the Hipparcos catalog. {The reason for this discrepancy is not clear to us.}

\section{Radial-velocities}
\label{sec:rv}

\begin{table*}[t]
\caption[]{Elements of the fitted orbits.}
\begin{tabular}{lllll}
\hline
\hline
\noalign{\smallskip}
			&  \object{HD\,5388} & \object{HD\,181720}	    & \object{HD190984}	& \\
\hline
$P$             	& 777$\pm$4             & 956$\pm$14			    & 4885$\pm$1600	& [d]\\
$T$             	& 2\,454\,570$\pm$9    & 2\,453\,631$\pm$30		    & 2\,449\,572$\pm$1600	& [d]\\
$a$			& 1.76                          & 1.78				    & 5.5                                 & [AU]\\
$e$             	& 0.40$\pm$0.02       & 0.26$\pm$0.06		    & 0.57$\pm$0.10           &  \\
$V_r$           	& 39.308$\pm$0.001& $-$45.3352$\pm$0.0004 & 20.269$\pm$0.004    & [km\,s$^{-1}$]\\
$\omega$        	& 324$\pm$4              & 177$\pm$12			   & 318$\pm$5 		& [degr] \\ 
$K_1$           	& 41.7$\pm$1.6          & 8.4$\pm$0.4			   & 48$\pm$1			& [m\,s$^{-1}$] \\
$f_1(m)$        	& 4.47\,10$^{-9}$	 &  0.053\,10$^{-9}$		   & 30.95				& [M$_{\odot}$]\\ 
$\sigma(O-C)$	& 3.33                           & 1.37				   & 3.44				& [m\,s$^{-1}$]  \\    
$N$             	& 68                              & 28					   & 47				&  \\
$m_2\,\sin{i}$  	& 1.96                           & 0.37				   & 3.1				 & [M$_{\mathrm{Jup}}$]\\
\noalign{\smallskip}
\hline
\end{tabular}
\label{tab:orbits}
\end{table*}

\subsection{HD\,5388}

Sixty-eight radial-velocity measurements of HD\,5388 were obtained between September 2003 and September 2009 (Table\,3)\footnote{Tables 3, 4, and 5, with the HARPS radial-velocities for HD\,5388, HD\,181720, and HD\,190984, respectively, are only available online.}.
The data show a clear periodic signal that can be well fitted with a Keplerian function with a period of 777\,days, an eccentricity of
0.40, and a semi-amplitude of 42\,m\,s$^{-1}$. This signal is compatible with the radial-velocity variation induced by a 1.96 Jupiter mass 
companion orbiting the 1.21M$_{\odot}$ dwarf HD\,5388 (see Table\,\ref{tab:orbits} and Fig.\,\ref{fig:HD5388_rv}). 

The analysis of the residuals of the Keplerian fit (3.33\,m\,s$^{-1}$) suggest that some residual trends may be present in the data.
However, we find no clear evidence for any  extra component. In any case, the average error of the individual data points 
is 2.8\,m\,s$^{-1}$, only 1.8\,m\,s$^{-1}$ below the residuals (after quadratic subtraction). Given the observational strategy 
used, the relatively high value for the stellar projected rotational velocity and the spectral type of the star, 
this value is likely not significant (see Sect.\,\ref{sec:sample}).

To understand if the periodic radial-velocity signal observed could have a non-planetary 
origin \citep[see e.g.][]{Saar-1997,Queloz-2000,Santos-2002a,Santos-2009}, we also analyzed the bisector inverse slope (BIS) 
of the cross-correlation Function \citep[][]{Queloz-2001}. No correlation between radial-velocity (RV) and BIS is seen (Fig.\,\ref{fig:bisvr}), which suggests that stellar activity
of stellar blends cannot explain the RV variation observed.
Together with the low activity level of the star, we conclude that the
777-day orbital period observed can be better explained by the presence of
a Jupiter-like planet which orbits \object{HD\,5388}.

\subsection{HD\,181720}

A total of 28 radial-velocity measurements of \object{HD\,181720} were taken between September 2003 and September 2009 (Table\,4). The velocities show a clear low amplitude and long period variation (Fig.\,\ref{fig:HD181720_rv}). The data are well fitted with a single Keplerian 
function with a period of 956\,days, an eccentricity of 0.26, and an amplitude 8.4\,m\,s$^{-1}$. Given the mass for HD\,181720, 
this signal can be explained by the presence of a 0.37 Jupiter-masses (minimum-mass) companion to HD\,181720 (Table\,\ref{tab:orbits}).

The residuals of the Keplerian fit (1.4\,m\,s$^{-1}$) are slighly higher than the average error of the individual radial-velocity measurements (1.0\,m\,s$^{-1}$).
A visual inspection of the residuals (Fig.\,\ref{fig:HD181720_rv}) suggests that some structure may exist in the data. However, the small number of points and 
the timeline of the measurements does not allow us to confirm this possibility.

The analysis of the bisector inverse slope shows that
no clear correlation exists between the BIS and the radial-velocities (Fig.\,\ref{fig:bisvr}).
Together with the low activity level of the star, we conclude that the
956-day orbital period observed can be better explained by the presence of
a Saturn-like planet which orbits \object{HD\,181720}.

\subsection{HD\,190984}

From June 2004 to September 2009 we obtained a total of 47 radial-velocity measurements HD\,190984 (Table\,5).
A look at the data (Fig.\,\ref{fig:HD190984_rv}) shows a clear long period radial-velocity signal. Though a complete
period is still not observable, the data can be well fitted with a Keplerian function with a period of 4885\,days, an eccentricity of 0.57, and a semi-amplitude of 48 m/s. This signal is expected from a 3.1 Jupiter mass companion orbiting HD\,190984.

The average photon noise error of the radial-velocities (1.7\,m\,s$^{-1}$) is significantly below the observed residuals (3.44\,m\,s$^{-1}$).
Given the stellar projected rotational velocity, spectral type, and evolutionary stage together with the observing 
strategy used (see Sect.\,\ref{sec:sample}),  this result is not unexpected. 
No correlation was found between BIS and the radial-velocities (Fig.\,\ref{fig:bisvr}). The observed radial-velocity signal is better explained as
due to the presence of a long period giant planet orbiting HD\,190984.

The fitted radial-velocity period is about twice as long as the baseline of our measurements. This increases the uncertainty of
the orbital solution, as can be clearly seen in the error bars in Table\,\ref{tab:orbits}. However, as preliminary as it can be,
it seems very unlikely that the observed signal is not due to the presence of a giant planet in orbit about HD\,190984. Since
the HARPS program which included this star is now over, we cannot assure that a correct follow-up will be done over
the next 5 years to better settle this result. We thus prefer to publish the present data, with a word of caution to say that the
listed orbital parameters may be subject to some adjustments.

\section{Concluding remarks}
\label{sec:conclusions}

We present the detection of three new giant planets orbiting three moderately metal-poor stars from two separate
HARPS GTO programs. This constitutes a strong addition to the previously known number of planets orbiting stars with metallicities 
significantly below solar.

\begin{figure}[t!]
\resizebox{8.5cm}{!}{\includegraphics{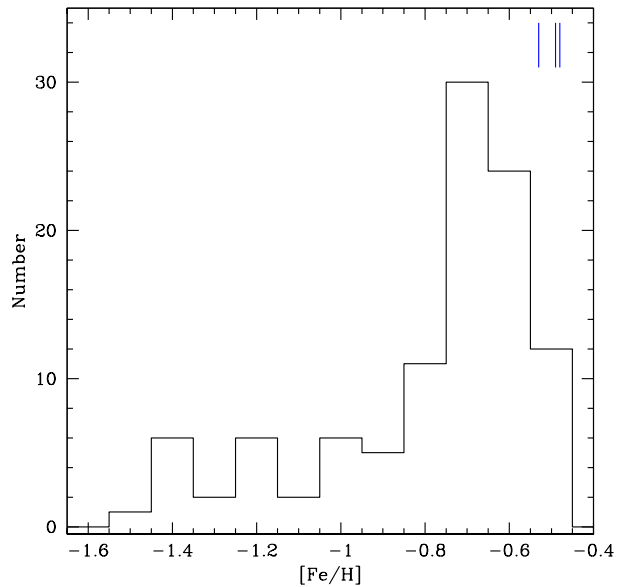}}
\caption{Metallicity distribution for the whole HARPS ``metal-poor'' sample. The three vertical lines denote the [Fe/H] of the three
stars in this sample for which a planet has been detected.}
\label{fig:histo}
\end{figure}

Two of these stars (HD\,181720 and HD\,190984) were part of a dedicated program to search for giant planets orbiting a sample of
$\sim$100 metal-deficient stars. Together with \object{HD\,171028}\,b \citep[][]{Santos-2007}, three planets have been 
discovered orbiting stars from this particular sample. In a very simplistic analysis, this suggests
that at least 3\% of stars with a metallicity below $\sim$$-$0.5\,dex have giant planets. Such a result tends to
confirm the hunch of \citet[][]{Santos-2004b} that the frequency distribution of giant planets as a function of stellar
metallicity is rather flat for [Fe/H] values below solar.
Interestingly,  HD\,171028, HD\,181720, and HD\,190984 all seem to be in the metal-rich tail of the metallicity distribution 
of the sample \citep[see][the former has a metallicity of $-$0.49\,dex]{Santos-2007}, which suggest that in this regime the frequency of planets is still 
an increasing function of the stellar metallicity (Fig.\,\ref{fig:histo}). 

It is also curious to see that all the three planets discovered in this sample have long period orbits {(above $\sim$1.5\,years)}, as do all the 
remaining planet candidates discovered that orbit solar-type dwarfs with [Fe/H] significantly below solar. No hot-Jupiter was found in 
the HARPS ``metal-poor'' sample.
The lack of short period planets in our metal-poor sample is also agrees with the results of the 
three-year long survey done by \citet[][]{Sozzetti-2009}. In their sample of 160 metal-poor dwarfs, no planet was discovered, though some 
long period trends were found. We note however, that Sozzetti et al. would likely have missed some of the planets in our sample (e.g. HD\,181720b, P$\sim$2.6\,years, K=8.4\,m\,s$^{-1}$) that induce RV semi-amplitudes smaller than the typical precision of their measurements ($\sim$10\,m\,s$^{-1}$).

A debate exists of whether there is a dependence between stellar metallicity and the orbital period of the discovered planets \citep[see discussion in][]{Santos-2006a}. Our results support this tendency. We note however, that a few short period transiting hot Jupiters have been found which orbit metal-poor stars \citep[see table in][]{Ammler-2009}.

These results may have important implications for the models of planet formation and evolution. A relatively high frequency of planets orbiting
metal-poor stars (in this case in long period orbits) could imply that the disk instability model is
at play at these low [Fe/H] values. Planet formation through disk instability is relatively insensitive to the metallicity of the disk (and of the host star). 
An increase in the number of known giant planets orbiting low metallicity stars is crucial to settle these issues.

\begin{acknowledgements}
The HARPS spectrograph has been build by the contributions of the Swiss FNRS,
the Geneva University, the French Institut National des Sciences de l'Univers (INSU) and ESO.
N.C.S. would like to thank the support by the European Research Council/European Community under the FP7 through a Starting Grant, as well as the support from Funda\c{c}\~ao para a Ci\^encia e a Tecnologia (FCT), Portugal, through program Ci\^encia\,2007. We would also like to acknowledge support from FCT in the form of grants reference PTDC/CTE-AST/098528/2008 and PTDC/CTE-AST/098604/2008. S.G.S and P.F. would like to acknowledge the support from the Funda\c{c}\~ao para a Ci\^encia e Tecnologia (Portugal) in the form of fellowships SFRH/BPD/47611/2008
and SFRH/BD/21502/2005, respectively. 

\end{acknowledgements}

\bibliographystyle{aa}
\bibliography{santos_bibliography}

\begin{thebibliography}{39}
\expandafter\ifx\csname natexlab\endcsname\relax\def\natexlab#1{#1}\fi

\bibitem[{{Ammler-von Eiff} {et~al.}(2009){Ammler-von Eiff}, {Santos}, {Sousa},
  {Fernandes}, {Guillot}, {Israelian}, {Mayor}, \& {Melo}}]{Ammler-2009}
{Ammler-von Eiff}, M., {Santos}, N.~C., {Sousa}, S.~G., {et~al.} 2009, ArXiv
  e-prints

\bibitem[{{Boss}(1997)}]{Boss-1997}
{Boss}, A.~P. 1997, Science, 276, 1836

\bibitem[{{Cochran} {et~al.}(2007){Cochran}, {Endl}, {Wittenmyer}, \&
  {Bean}}]{Cochran-2007}
{Cochran}, D.~C., {Endl}, M., {Wittenmyer}, R.~A., \& {Bean}, J.~L. 2007, ApJ,
  in press

\bibitem[{{Dodson-Robinson} {et~al.}(2009){Dodson-Robinson}, {Veras}, {Ford},
  \& {Beichman}}]{DodsonRobinson-2009}
{Dodson-Robinson}, S.~E., {Veras}, D., {Ford}, E.~B., \& {Beichman}, C.~A.
  2009, ArXiv e-prints

\bibitem[{{ESA}(1997)}]{ESA-1997}
{ESA}. 1997, The Hipparcos and Tycho Catalogues

\bibitem[{{Fernandes} \& {Santos}(2004)}]{Fernandes-2004}
{Fernandes}, J. \& {Santos}, N.~C. 2004, A\&A, 427, 607

\bibitem[{{Fischer} \& {Valenti}(2005)}]{Fischer-2005}
{Fischer}, D.~A. \& {Valenti}, J. 2005, ApJ, 622, 1102

\bibitem[{{Flower}(1996)}]{Flower-1996}
{Flower}, P.~J. 1996, ApJ, 469, 355

\bibitem[{{Girardi} {et~al.}(2000){Girardi}, {Bressan}, {Bertelli}, \&
  {Chiosi}}]{Girardi-2000}
{Girardi}, L., {Bressan}, A., {Bertelli}, G., \& {Chiosi}, C. 2000, A\&AS, 141,
  371

\bibitem[{{Gonzalez}(1997)}]{Gonzalez-1997}
{Gonzalez}, G. 1997, MNRAS, 285, 403

\bibitem[{{Henry} {et~al.}(1996){Henry}, {Soderblom}, {Donahue}, \&
  {Baliunas}}]{Henry-1996}
{Henry}, T.~J., {Soderblom}, D.~R., {Donahue}, R.~A., \& {Baliunas}, S.~L.
  1996, AJ, 111, 439

\bibitem[{{Ida} \& {Lin}(2004)}]{Ida-2004a}
{Ida}, S. \& {Lin}, D.~N.~C. 2004, ApJ, 604, 388

\bibitem[{{Kalas} {et~al.}(2008){Kalas}, {Graham}, {Chiang}, {Fitzgerald},
  {Clampin}, {Kite}, {Stapelfeldt}, {Marois}, \& {Krist}}]{Kalas-2008}
{Kalas}, P., {Graham}, J.~R., {Chiang}, E., {et~al.} 2008, Science, 322, 1345

\bibitem[{{Marois} {et~al.}(2008){Marois}, {Macintosh}, {Barman}, {Zuckerman},
  {Song}, {Patience}, {Lafreni{\`e}re}, \& {Doyon}}]{Marois-2008}
{Marois}, C., {Macintosh}, B., {Barman}, T., {et~al.} 2008, Science, 322, 1348

\bibitem[{{Masana} {et~al.}(2006){Masana}, {Jordi}, \& {Ribas}}]{Masana-2006}
{Masana}, E., {Jordi}, C., \& {Ribas}, I. 2006, A\&A, 450, 735

\bibitem[{{Matsuo} {et~al.}(2007){Matsuo}, {Shibai}, {Ootsubo}, \&
  {Tamura}}]{Matsuo-2007}
{Matsuo}, T., {Shibai}, H., {Ootsubo}, T., \& {Tamura}, M. 2007, ApJ, 662, 1282

\bibitem[{{Mayor} {et~al.}(2003){Mayor}, {Pepe}, {Queloz}, {Bouchy},
  {Rupprecht}, {Lo Curto}, {Avila}, {Benz}, {Bertaux}, {Bonfils}, {dall},
  {Dekker}, {Delabre}, {Eckert}, {Fleury}, {Gilliotte}, {Gojak}, {Guzman},
  {Kohler}, {Lizon}, {Longinotti}, {Lovis}, {Megevand}, {Pasquini}, {Reyes},
  {Sivan}, {Sosnowska}, {Soto}, {Udry}, {van Kesteren}, {Weber}, \&
  {Weilenmann}}]{Mayor-2003b}
{Mayor}, M., {Pepe}, F., {Queloz}, D., {et~al.} 2003, The Messenger, 114, 20

\bibitem[{{Mordasini} {et~al.}(2009){Mordasini}, {Alibert}, \&
  {Benz}}]{Mordasini-2009a}
{Mordasini}, C., {Alibert}, Y., \& {Benz}, W. 2009, A\&A, 501, 1139

\bibitem[{{Naef} {et~al.}(2007){Naef}, {Mayor}, {Benz}, {Bouchy}, {Lo Curto},
  {Lovis}, {Moutou}, {Pepe}, {Queloz}, {Santos}, \& {Udry}}]{Naef-2007}
{Naef}, D., {Mayor}, M., {Benz}, W., {et~al.} 2007, A\&A, 470, 721

\bibitem[{{Nordstr{\"o}m} {et~al.}(2004){Nordstr{\"o}m}, {Mayor}, {Andersen},
  {Holmberg}, {Pont}, {J{\o}rgensen}, {Olsen}, {Udry}, \&
  {Mowlavi}}]{Nordstrom-2004}
{Nordstr{\"o}m}, B., {Mayor}, M., {Andersen}, J., {et~al.} 2004, A\&A, 418, 989

\bibitem[{{Pepe} {et~al.}(2004){Pepe}, {Mayor}, {Queloz}, {Benz}, {Bonfils},
  {Bouchy}, {Curto}, {Lovis}, {M{\' e}gevand}, {Moutou}, {Naef}, {Rupprecht},
  {Santos}, {Sivan}, {Sosnowska}, \& {Udry}}]{Pepe-2004}
{Pepe}, F., {Mayor}, M., {Queloz}, D., {et~al.} 2004, A\&A, 423, 385

\bibitem[{{Pollack} {et~al.}(1996){Pollack}, {Hubickyj}, {Bodenheimer},
  {Lissauer}, {Podolak}, \& {Greenzweig}}]{Pollack-1996}
{Pollack}, J., {Hubickyj}, O., {Bodenheimer}, P., {et~al.} 1996, Icarus, 124,
  62

\bibitem[{{Queloz} {et~al.}(2001){Queloz}, {Henry}, {Sivan}, {Baliunas},
  {Beuzit}, {Donahue}, {Mayor}, {Naef}, {Perrier}, \& {Udry}}]{Queloz-2001}
{Queloz}, D., {Henry}, G.~W., {Sivan}, J.~P., {et~al.} 2001, A\&A, 379, 279

\bibitem[{{Queloz} {et~al.}(2000){Queloz}, {Mayor}, {Weber}, {Bl{\' e}cha},
  {Burnet}, {Confino}, {Naef}, {Pepe}, {Santos}, \& {Udry}}]{Queloz-2000}
{Queloz}, D., {Mayor}, M., {Weber}, L., {et~al.} 2000, A\&A, 354, 99

\bibitem[{{Saar} \& {Donahue}(1997)}]{Saar-1997}
{Saar}, S.~H. \& {Donahue}, R.~A. 1997, ApJ, 485, 319

\bibitem[{{Santos} {et~al.}(2004{\natexlab{a}}){Santos}, {Bouchy}, {Mayor},
  {Pepe}, {Queloz}, {Udry}, {Lovis}, {Bazot}, {Benz}, {Bertaux}, {Lo Curto},
  {Delfosse}, {Mordasini}, {Naef}, {Sivan}, \& {Vauclair}}]{Santos-2004a}
{Santos}, N.~C., {Bouchy}, F., {Mayor}, M., {et~al.} 2004{\natexlab{a}}, A\&A,
  426, L19

\bibitem[{{Santos} {et~al.}(2004{\natexlab{b}}){Santos}, {Israelian}, \&
  {Mayor}}]{Santos-2004b}
{Santos}, N.~C., {Israelian}, G., \& {Mayor}, M. 2004{\natexlab{b}}, A\&A, 415,
  1153

\bibitem[{{Santos} {et~al.}(2007){Santos}, {Mayor}, {Bouchy}, {Pepe}, {Queloz},
  \& {Udry}}]{Santos-2007}
{Santos}, N.~C., {Mayor}, M., {Bouchy}, F., {et~al.} 2007, A\&A, 474, 647

\bibitem[{{Santos} {et~al.}(2000){Santos}, {Mayor}, {Naef}, {Pepe}, {Queloz},
  {Udry}, \& {Blecha}}]{Santos-2000a}
{Santos}, N.~C., {Mayor}, M., {Naef}, D., {et~al.} 2000, A\&A, 361, 265

\bibitem[{{Santos} {et~al.}(2002){Santos}, {Mayor}, {Naef}, {Pepe}, {Queloz},
  {Udry}, {Burnet}, {Clausen}, {Helt}, {Olsen}, \& {Pritchard}}]{Santos-2002a}
{Santos}, N.~C., {Mayor}, M., {Naef}, D., {et~al.} 2002, A\&A, 392, 215

\bibitem[{{Santos} {et~al.}(2006){Santos}, {Pont}, {Melo}, {Israelian},
  {Bouchy}, {Mayor}, {Moutou}, {Queloz}, {Udry}, \& {Guillot}}]{Santos-2006a}
{Santos}, N.~C., {Pont}, F., {Melo}, C., {et~al.} 2006, A\&A, 450, 825

\bibitem[{{Santos} {et~al.}(2009){Santos}, {Silva}, {Lovis}, \&
  {Melo}}]{Santos-2009}
{Santos}, N.~C., {Silva}, J., {Lovis}, C., \& {Melo}, C. 2009, A\&A, submitted

\bibitem[{{Sousa} {et~al.}(2008){Sousa}, {Santos}, {Mayor}, {Udry},
  {Casagrande}, {Israelian}, {Pepe}, {Queloz}, \& {Monteiro}}]{Sousa-2008}
{Sousa}, S.~G., {Santos}, N.~C., {Mayor}, M., {et~al.} 2008, A\&A, 487, 373

\bibitem[{{Sozzetti} {et~al.}(2006){Sozzetti}, {Torres}, {Latham}, {Carney},
  {Stefanik}, {Boss}, {Laird}, \& {Korzennik}}]{Sozzetti-2006b}
{Sozzetti}, A., {Torres}, G., {Latham}, D.~W., {et~al.} 2006, ApJ, 649, 428

\bibitem[{{Sozzetti} {et~al.}(2009){Sozzetti}, {Torres}, {Latham}, {Stefanik},
  {Korzennik}, {Boss}, {Carney}, \& {Laird}}]{Sozzetti-2009}
{Sozzetti}, A., {Torres}, G., {Latham}, D.~W., {et~al.} 2009, \apj, 697, 544

\bibitem[{{Udry} {et~al.}(2000){Udry}, {Mayor}, {Naef}, {Pepe}, {Queloz},
  {Santos}, {Burnet}, {Confino}, \& {Melo}}]{Udry-2000}
{Udry}, S., {Mayor}, M., {Naef}, D., {et~al.} 2000, A\&A, 356, 590

\bibitem[{{Udry} \& {Santos}(2007)}]{Udry-2007}
{Udry}, S. \& {Santos}, N. 2007, ARAA, 45, 397

\bibitem[{{van Leeuwen}(2007)}]{vanLeeuwen-2007}
{van Leeuwen}, F., ed. 2007, Astrophysics and Space Science Library, Vol. 350,
  {Hipparcos, the New Reduction of the Raw Data}

\bibitem[{{Wright}(2004)}]{Wright-2004}
{Wright}, J.~T. 2004, AJ, 128, 1273

\end{thebibliography}

\end{document}